# Ramadan and Infants Health Outcomes

## Hossein Abbaszadeh Shahri[1]

### Abstract

Previous studies show that prenatal shocks to embryos could have adverse impacts on health endowment at birth. Using the universe of birth data and a difference-in-difference-in-difference strategy, I find that exposure to Ramadan during prenatal development has negative birth outcomes. Exposure to a full month of fasting is associated with 96 grams lower birth-weight. These results are robust across specifications and do not appear to be driven by mothers' selective fertility.

[1] Department of Environment, Texas Tech University, 2500 Broadway, Lubbock, TX 79409; ORCID ID: https://orcid.org/0000-0002-3435-9914, Email: hossein.shahri@ttu.edu; Phone: +1-409-454-7711

# 1. Introduction

Nutrition is the main contributor to the environmental elements that generate the fetal genome. A nutritional distortion may lead the fetus to adopt in a way that alters the structure and physiology of the newborn, which, in turn, affects later-life outcomes. This process of self-adaptation and so-called *fetal programming* is the basis of the theory of *Fetal Origin* (Almond and Currie, 2011; Wu, Bazer, Cudd, Meininger, and Spencer, 2004). Investigating the fetal roots of later life outcomes has recently captured attention among health economists (Behrman and Rosenzweig, 2004; Hoynes et al., 2011; Myrskylä, 2010; NoghaniBehambari et al., 2020a, 2020b; Sorensen et al., 1999; Tavassoli et al., 2020). For instance, (Hoynes et al., 2015) explore the effect of reforms of Earned Income Tax Credit as a source of permanent shock to income of the disadvantaged population and find that the improvements in the welfare had potential to improve birth outcomes.

A small strand of this literature evaluates the effect of Ramadan observance during antenatal development as a form of gentle malnutrition (Almond, Mazumder, and Van Ewijk, 2014; Jürges, 2015; Majid, 2015). Using Michigan natality detailed files (1989-2006) and a difference-in-difference strategy, Almond and Mazumder (2011) (hereafter AM) find that exposure to a full month of Ramadan among Arab mothers (versus non-Arabs) in Michigan during pregnancy is associated with, on average, 18 grams lower birth-weight and roughly 6 percentage point lower fraction of male births. The reduction in birth-weight is more pronounced during first trimester (20 grams) and second trimester (25 grams).

I improve their findings using US natality detailed files over the years 2004-2017 and applying a difference-in-difference-in-difference identification strategy. I find that a full-month exposure to Ramadan reduces the birth-weight by 80-110 grams. The effects are larger than those of AM by a factor of 3 to 4. There are three potential drivers for the observed difference. First, they recognize Muslims based on the race (Arabs VS non-Arabs). Since the current data asks for mother's birth country, I can distinguish the religion using aggregate distribution of Muslims in home-country. In fact, when I restrict the sample to Arab mothers in Michigan the estimated coefficients are very close to those of AM. Second, Arabs in Michigan are highly geographically gathered[2]. Using all Muslim mothers in Michigan will provide coefficients two-times larger than those of Arab mothers. Third, there is very limited variation of day-light hours across geographic locations in AM. The nationwide nature of the current data provides a much larger dosage of day-time exposure. Daylight hours vary from about 9.6 hours (Alaska, 2004) to 15.9 hours (Ohio, 2017). Ramadan occurs at lunar calendar and moves back about ten days each year. It allows for a source of variation of exposure over time. The across-state and over-time variation of the treatment in the data is much deeper than AM and provides more refined estimates.

---
[2] About 80% of Arabs in Michigan reside in only three counties.

Similar to AM, the main drawback of the data is that it lacks information about whether mothers did actually observe Ramadan. Therefore, the estimates must be interpreted as an intention-to-treat effect and a minimum causal effect. However, some researchers have found that 50-90% of pregnant Muslim mothers do observe Ramadan (e.g., Joosoph, Abu, Yu, and others, 2004; Mubeen, Mansoor, Hussain, and Qadir, 2012; and for US: Robinson and Raisler, 2005).

## 2. Data, Sample Selection and Empirical Method

The data are extracted from US Natality Detailed files 2004-2017. The sample is restricted to singleton birth records who reach full-term gestational age. I also restrict the sample to mothers who were definitely exposed to a full month of Ramadan and those who were surely not exposed. I use OLS regressions of the following form:

$$y_{itmc} = \alpha + \sum_{k=1}^{3} \beta^k \text{ExpHours}_{tmc}^k + \pi X_{itmc} + \emptyset Z_{ct} + \mu_t + \gamma_m + \rho_c + \epsilon_{itmc} \quad (1)$$

Hours of exposure to Ramadan, $ExpHours$, vary over each trimester $k$, the birth years $t$, month of gestation $m$, and also based on the latitude of the county $c$. Some mother and father characteristics are included in $X$. In $Z$, I include some county-by-year characteristics. The set of $\beta^k$ are the coefficients of interest that capture the effect of hours of exposure on the outcome. I also report the results for a continuous variable that indicates the total hours of exposure. Hourly exposures are divided by the average Ramadan hours over the sample period.

## 3. Results

A summary statistics of the final sample is reported in Table 1. On average, birth weight of infants is 3,327 grams. The main results of regressions introduced in equation 1 are reported in

Table 2. The first column focuses on AM's sample. The estimates are quite close to their findings. In column 2, I restrict the sample to Michigan-resident immigrant mothers from countries in which Muslims' share exceeds 90%. The coefficients are almost two times larger. In column 3, the results for all US Muslims are reported. A full month exposure is associated with 138 grams fewer birth-weight. I use two non-Muslim groups as potential control groups: immigrants from non-Muslim countries (column 4) and US-born whites (column 5). There is no significant effect of exposure on their birth outcome. In order to rule out the seasonality effects, I apply a difference-in-difference strategy using non-Muslim immigrants as control group (column 6) and US-born mothers as control group (column 7). While the results of the latter control group is very close to that of Muslim samples only (128 grams reduction in birth-weight) the coefficients of DD using the former control group is smaller (75 grams reduction). One concern is that the seasonality effects act different for immigrants (Muslim and non-Muslims) compared to US-born mothers. To address this issue, I use a difference-in-difference-in-difference strategy by interacting an indicator of being immigrant and an indicator of being Muslim to all right-hand-side variables except the county dummies. The results are reported in column 9. A full month of exposure during first, second, and third trimester is associated with 82.3, 79.8, and 109.5 grams lower birth-weight, respectively. These effects are much larger than AM but comparable to findings of Savitri et al. (2014) who find that Ramadan fasting among Muslim mothers reduces birth-weight by about 272 grams. The effects are also similar to findings of Haeck and Lefebvre (2016) who investigate the effect of egg-milk-orange program, a nutritional program for pregnant women, on birth outcomes. They find that the nutritional support could increase the birth-weight by 70 grams.

As a falsification test, I assign US-born mothers a fake Muslim status and non-Muslim immigrants serve as the control group. The DD results are reported in column 8. None of the coefficients is statistically significant and economically large.

Another concern is that mothers might target their pregnancy timing to avoid any overlap with Ramadan. If some characteristics of mothers, like education, make them more health conscious to avoid untimely pregnancy and meanwhile these characteristics affect the birth-weight, then the OLS results are biased. I try to address this issue by using mother's education, a proxy of mother's characteristics and socioeconomic status, as the outcome in equation 1. Results of single sub-sample, DiD, and DiDiD strategy (reported in Table 3) rule out this possibility. Although the coefficients on Muslim mothers are relatively large, they are positive and insignificant.

## 4. Conclusion

Using a sample of over 16M births from US natality files 2004-2017 and a DiDiD strategy, I find that exposure to a full month of fasting among Muslim mothers is associated with, on average, 96 grams lower birth-weight. This is in line with previous literature but the point estimates are larger than those found by AM. The DiDiD strategy and the falsification tests rule out the possibility that my findings are driven by seasonality effects or the fact that the time trend effects act differently for immigrant mothers versus US-born mothers.

Overall, the estimates are intention-to-treat effects in a reduced form analysis. Not having information about the first stage effects, one might interpret the coefficients with caution since the real causal effects might be larger. If only 75% of Muslim mothers did observe Ramadan (a number within empirical range of 50-90%), the results imply a negative effect of 128 grams, or a 3.7% decrease from the mean birth-weight among Muslims.

# Tables

### Table 1 - Summary Statistics

| Variable | Observations | Mean | Std. Dev. | Min | Max |
|---|---|---|---|---|---|
| *Infant Characteristics:* | | | | | |
| Birth Weight (grams) | 16,547,046 | 3327.982 | 602.795 | 227 | 8165 |
| Gestational Weeks | 16,547,046 | 39.043 | 2.700 | 17 | 52 |
| Sex (f=1) | 16,547,046 | 0.488 | 0.497 | 0 | 1 |
| *Mother Characteristics:* | | | | | |
| Age | 16,547,046 | 26.465 | 5.886 | 10 | 54 |
| Race: White | 16,547,046 | 0.796 | 0.402 | 0 | 1 |
| Race: Black | 16,547,046 | 0.160 | 0.367 | 0 | 1 |
| Unmarried | 16,547,046 | 0.283 | 0.450 | 0 | 1 |
| Education (Years of Schooling) | 16,547,046 | 12.625 | 2.654 | 0 | 17 |
| Month Prenatal Care Began | 16,547,046 | 2.596 | 1.517 | 0 | 9 |
| Prenatal Visits | 16,547,046 | 11.179 | 4.025 | 0 | 49 |
| *State Characteristics:* | | | | | |
| GSP per capita | 16,547,046 | 43585.268 | 9031.635 | 24371.631 | 140143.05 |
| Personal Income per capita | 16,547,046 | 371.483 | 66.910 | 212.533 | 624.262 |
| %Blacks | 16,547,046 | 12.653 | 8.174 | .222 | 69.376 |
| %Whites | 16,547,046 | 83.354 | 8.514 | 27.002 | 99.301 |
| %Males | 16,547,046 | 48.827 | 0.709 | 46.263 | 53.005 |
| %Population 25-65 | 16,547,046 | 50.716 | 2.344 | 40.368 | 55.143 |
| Log Current Transfer Receipt | 16,547,046 | 18.080 | 0.991 | 14.495 | 19.850 |
| Log Income Maintenance Benefits | 16,547,046 | 15.830 | 1.131 | 11.503 | 17.908 |
| Log Unemployment Insurance Benefits | 16,547,046 | 14.594 | 1.119 | 10.697 | 16.796 |
| Log Other Welfare Benefits | 16,547,046 | 17.923 | 0.978 | 14.056 | 19.657 |
| Minimum Wage | 16,547,046 | 7.481 | 0.813 | 6.266 | 11.409 |

Notes. The data covers the years 2004-2017. All dollar values are converted into 2000 dollars to reflect real values.

## Table 2- Prenatal Ramadan Exposure and Birth Outcomes

|  | Michigan Arabs | Michigan Muslims | All US Muslims | Immigrants Non-Muslims | US Born Whites | DD (3-4) | DD (3-5) | DD (4-5) | DDD (3-4-5) |
|---|---|---|---|---|---|---|---|---|---|
|  | (1) | (2) | (3) | (4) | (5) | (6) | (7) | (8) | (9) |
| *Outcome: Birth-weight* | | | | | | | | | |
| Ramadan Hours | -24.62* | -55.15*** | -138.23** | -10.30 | 2.53 | -74.75** | -127.56*** | 5.43 | -96.03*** |
|  | (14.05) | (14.38) | (62.25) | (12.41) | (5.84) | (32.41) | (32.35) | (11.75) | (32.78) |
| $R^2$ | 0.07 | 0.08 | 0.08 | 0.05 | 0.06 | 0.05 | 0.06 | 0.06 | 0.06 |
| Observations | 17,603 | 16,866 | 215,815 | 3,983,943 | 12,347,288 | 4,199,758 | 12,563,103 | 16,331,231 | 16,547,046 |
| Mean DV | 3390.54 | 3351.63 | 3388.45 | 3434.24 | 3483.25 | 3431.86 | 3481.59 | 3471.24 | 3470.14 |
| *Ramadan Hours During:* | | | | | | | | | |
| First Trimester | -13.42 | -37.28** | -105.89* | -11.39 | 1.94 | -62.52** | -113.56*** | 3.47 | -82.27*** |
|  | (14.62) | (14.62) | (54.51) | (11.23) | (6.14) | (29.74) | (30.15) | (10.99) | (30.55) |
| Second Trimester | -25.36* | -47.55*** | -101.68** | -8.48 | 5.50 | -62.94** | -105.02*** | -5.87 | -79.86*** |
|  | (13.52) | (13.92) | (51.37) | (10.52) | (5.58) | (27.89) | (27.45) | (11.25) | (27.81) |
| Third Trimester | -35.98** | -77.41*** | -141.25* | -13.95 | 3.60 | -85.36** | -147.89*** | 8.81 | -109.55*** |
|  | (17.89) | (18.39) | (71.95) | (14.71) | (7.40) | (39.01) | (38.89) | (14.72) | (39.28) |
| $R^2$ | 0.07 | 0.08 | 0.08 | 0.05 | 0.06 | 0.05 | 0.06 | 0.06 | 0.06 |
| Observations | 17,603 | 16,866 | 215,815 | 3,983,943 | 12,347,288 | 4,199,758 | 12,563,103 | 16,331,231 | 16,547,046 |
| Mean DV | 3390.54 | 3351.63 | 3388.45 | 3434.24 | 3483.25 | 3431.86 | 3481.59 | 3471.24 | 3470.14 |

Notes: all regressions include county characteristics (real per capita income, percentage whites, percentage blacks, percentage male, percentage manufacturing, and real wages), dummied for mother's characteristics (race, education, age, marital status, Hispanic origin, birth order, and cigarette smoking), dummies for father's characteristics (age, Hispanic origin, and race), and dummies for missing indicator of mother's and father's characteristics. All regressions also include county, month of gestation, year of birth, and mother's country of origin fixed effects. Standard errors are reported in parentheses.

\* $p < 0.1$
\*\* $p < 0.05$

**Table 3 – Selective Fertility: Ramadan and Mothers' Characteristics**

|  | Outcome: Education | Outcome: Marital Status |
|---|:---:|:---:|
|  | (1) | (2) |
| *Exposure dummy* |  |  |
|  | -0.73 | -0.25 |
|  | (1.79) | (1.58) |
| $R^2$ | 0.29 | 0.01 |
|  |  |  |
| *Daylight Hours of Exposure* |  |  |
|  | -0.0045 | 0.0001 |
|  | (0.0046) | (0.0002) |
| $R^2$ | 0.29 | 0.001 |
|  |  |  |
| *Daylight Hours of Exposure by Trimester* |  |  |
| First Trimester | -0.002 | -0.001 |
|  | (0.005) | (0.005) |
| Second Trimester | -0.002 | -0.003 |
|  | (0.005) | (0.004) |
| Third Trimester | -0.006 | 0.003 |
|  | (0.005) | (0.004) |
| $R^2$ | 0.29 | 0.001 |
|  |  |  |
| Observations | 15,044,225 | 15,051,455 |

Notes. Robust standard errors, clustered at county level, are reported in parentheses. Individual and county covariates are explained in the text.
* $p < 0.1$
** $p < 0.05$